\documentclass[twocolumn,showpacs,preprintnumbers,amsmath,amssymb]{revtex4}
\usepackage{graphicx}
\usepackage{dcolumn}
\usepackage{bm}
\begin{document}

\title{Lifshitz transitions and zero point lattice fluctuations in sulfur hydride showing near room temperature superconductivity}

\author{Antonio Bianconi$^{1,2,3}$ and Thomas Jarlborg$^4$}

\affiliation{
$^1$ RICMASS, Rome International Center for Materials Science Superstripes, Via dei Sabelli 119A, 00185 Rome, Italy
\\
$^2$ Institute of Crystallography, Consiglio Nazionale delle Ricerche, via Salaria, 00015 Monterotondo, Italy
\\
$^3$ INSTM, Consorzio Interuniversitario Nazionale per la Scienza e Tecnologia dei Materiali, Rome Udr, Italy
\\
$^4$DPMC, University of Geneva, 24 Quai Ernest-Ansermet, CH-1211 Geneva 4, Switzerland}


\begin{abstract}

Emerets's experiments on pressurized sulfur hydride have shown 
that H$_3$S metal has the highest known superconducting critical temperature 
$T_c=203$K. The Emerets data show pressure induced changes of the isotope coefficient 
between 0.25 and 0.5, in disagreement with Eliashberg
 theory which predicts a nearly constant isotope coefficient. We assign the pressure dependent 
isotope coefficient to Lifshitz  transitions induced by pressure and zero point lattice fluctuations. It is known that pressure could induce changes of the topology 
 of the Fermi surface, called Lifshitz transitions, but were neglected in previous papers 
 on the H$_3$S superconductivity issue. 
 Here we propose that H$_3$S is a multi-gap superconductor with a first condensate
 in the BCS regime (in the large Fermi surface with high Fermi energy) which coexists with 
 a second condensates in the BCS-BEC crossover regime (located on a small Fermi surface 
 spots with small Fermi energy) near the $\Gamma$ and M point. We discuss the need of Bianconi-Perali-Valletta (BPV) superconductivity theory 
for superconductivity in H$_3$S. It includes both the correction of the chemical potential 
due to pairing and the configuration interaction between different condensates, neglected 
by the Eliashberg  theory. Here the shape resonance in superconducting gaps, similar to
 Feshbach resonance in ultracold gases, gives a relevant contribution to amplify the critical temperature.
  Therefore this work provides some key tools needed in the search for new room temperature superconductors.

\end{abstract}


\maketitle

\section{Introduction.}

Following early claims of 190 K superconductivity in sulfur
 hydride at very high pressure \cite{droz}, new results of near room 
 temperature superconductivity with T$_c$=203 K i.e., at only -70 $^o$C \cite{droz2} have been presented
 on June 17, 2015 at Superstripes 2015 conference in Ischia, 
 Italy \cite{eremets} and they have triggered a very high scientific  interest \cite{carlidge}. The recent work of Emerets's 
group \cite{droz2} shows the Meissner effect and the pressure dependent critical temperature of H$_3$S and D$_3$S. 
These results have triggered today the materials research
 for room temperature superconductors in different hydrides at extreme high pressures  \cite{Ma2,Chen,Hou,Pau,Liu}. 

The experimental discovery of near room temperature 
superconductivity in sulfur hydride H$_2$S at very high pressure was 
predicted by \cite{duan} to occur in the high pressure metallic H$_3$S  phase, with $Im\bar{3}m$ lattice symmetry. 
Disproportion from 2(H$_2$S) + H$_2$  to 2(H$_3$S) occurs at very high pressure. 
The theoretical prediction of Duan et al. \cite{duan} has been obtained by 
using the successful theoretical approach used to predict crystalline structures 
at high pressure in material science: the evolutionary algorithm Universal Structure Predictor: Evolutionary Xtallography (USPEX)  \cite{oganov}. 
 
\begin{figure}
\includegraphics[height=8.0cm,width=5.0cm]{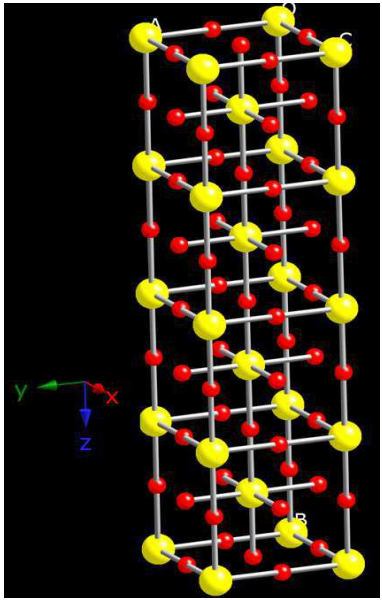}
\caption{(Color online) The crystalline structure of H$_3$S with
 $Im\bar{3}m$ lattice symmetry, with two formula units per unit cell, 
 with sulfur (yellow large spheres) and hydrogen atoms (red small spheres). 
 The linear S-H-S  hydrogen bonds along the a,b and c axis in the $Im\bar{3}m$ 
 lattice of  H$_3$S, form a first 3D network of three linear chains of covalent bonds 
 crossing at the sulfur atom at S1(0,0,0,) which coexists with  a second 3D network 
 of other three linear chains of covalent bonds (solid black lines) crossing at the sulfur atom at S1(0.5,0.5,0.5).}
\label{Im3ma}
\end{figure}

Many of preceding theoretical studies on this issue conclude that the superconducting phase 
in pressurized H$_3$S is described by the Eliashberg
theory \cite{duan,duan2,sanna,ma1, papa,errea,bernstein, durajski,ari}  while Hirsh 
proposes the hole superconductivity model  \cite{marsiglio}.

The Eremets's group research was motivated by the search 
for room temperature superconductivity predicted to emerge in metallic 
hydrogen and hydrides \cite{ash1,Ginz,ash2,mak,ash3,ash4,ash5,mak2,Abe}. 

The BCS theory \cite{BCS} has given a microscopic description
 of the superconducting condensate many body wave-function made 
 of interacting Cooper pairs in a weak coupling regime, where the pairing is mediated 
 by the conventional attractive phonon-exchange mechanism. We call here standard BCS theory
 the BCS formulas \cite{BCS} obtained with many approximations, 
 valid for a simple homogeneous crystal, with a single band and isotropic
  pairing, using a single value of the density of states at the Fermi level $N_0$ and a constant 
  coupling constant $\lambda$. Moreover the standard BCS theory assumes 
 a very small energy of the phonon and a very high electron density
  i.e., a high Fermi energy $ \omega_0/E_F <<1$ called Migdal approximation \cite{mig} 
  within the adiabatic Born-Oppenheimer approximation, where the electronic 
  and ionic degrees of freedom can be rigorously separated. The prediction 
  of the superconducting critical temperature in the frame of the standard 
  BCS theory approximations has required the introduction of the electron-electron 
  repulsive interaction and corrections due to strong coupling given by 
  the McMillan \cite{mcm,dyn} and the Eliashberg \cite{elia} formula. 

It was rapidly well accepted that the critical temperature cannot be 
larger than 30 K in the frame of standard BCS theory based only on 
the role of high energy phonons and strong electron-phonon coupling \cite{Ginz,mak,mak2}. 
In fact in the single band approximation $T_c$ increases with both phonon energy 
and coupling strength but for extreme strong electron-phonon coupling the electron 
liquid at low temperature prefers to order in the real space, forming electronic 
crystals like, charge density waves, spin density waves, which compete with 
the superconducting phase. Moreover if superconductivity survives increasing 
electron-phonon coupling, the critical temperature  decreases since the phonon 
energy is pushed toward zero. In these regime the lattice structure collapses and 
the system is in the verge of a catastrophe. Therefore it was known that materials 
research for room temperate superconductivity cannot be driven simply by looking 
to increase the phonon energy and coupling strength in the cooper pairing. 
The theoretical predictions of high $T_c$ in solid hydrogen and hydrates at 
high pressure were based not only on the high frequency phonons 
mediating the pairing in solid hydrogen or hydrates but also
 on joint control of the electron electron interaction via the changes of the dielectric constant 
 controlling the repulsive electron-electron interaction and the Coulomb 
 energy \cite{ash1,Ginz,ash2,mak,ash3,ash4,ash5,mak2,Abe} which can 
 become negative  \cite{Ginz,mak,mak2}  for an electronic system in the low density limit. 
 
The self consistent quantum many body theory of superconductivity, avoiding
 the BCS approximations, was made by Bogoliubov \cite{bog}, giving the fundamental
 spectrum of excited quasiparticles. It was developed by 
 Gorkov \cite{gor} and  Blatt \cite{bla} including the contribution to 
 the critical temperature of the condensation energy, related with 
 Josephson-like \cite{jos} terms, \cite{ann,bab} and considering multiple 
 symmetries of possible multiple condensates with a single critical temperature \cite{leg}. 
 A second route to rise the critical temperature of superconductivity
toward the maximum possible energy,  $K_BT_c /E_F\sim1$  was proposed 
in 1994 based on the non standard BCS theories where the control of $T_c$
can be achieved not only by increasing the pairing strength but also 
 by the control of the condensation energy, via a pair exchange mechanism between condensates, 
called the shape resonance in superconducting gaps \cite{cuprates1,cuprates11,cuprates12,cuprates13}. 
It was proposed that cuprates are actually complex inhomogeneous
materials with lattice quenched disorder, lattice modulations and short 
range charge density waves giving multiple Fermi surface arcs where
in each Fermi arc there is a different superconducting condensate. 
 One of these condensates is in the polaronic regime, where the Migdal 
approximation is violated $\omega_0 /E_F\sim1$, and the condensate is 
in the BCS-BEC crossover regime while the other condensates are within
the Migdal approximation  $ \omega_0/E_F <<1$. 
In 1996-1998 \cite{shape1,shape11,shape12,isotope1,shape} a superconductivity theory was developed by Bianconi,
Perali, Valletta (BPV) based on the Blatt \cite{bla} and Legget  \cite{leg} theories. 
The superconducting critical temperature here is not only controlled only by
the cooper pair formation but also by the exchange terms between 
pairs. These terms control the condensation energy and phase coherence and are
unavoidable ingredients for the formation and stability of high temperature superconducting 
condensate with short coherence length. 
A similar mechanism, called Feshbach resonance, was proposed later to increase the critical temperature
 for the formation of the superfluid condensate in ultracold  atoms \cite{fes1,fes2}.
Later other many body theories have been proposed aimed to control 
the condensation energy and the global phase coherence in superconductors and superfluids \cite{fes3,fes4}. 

Details of the normal metallic phase 
become key ingredients in these non standard BCS theories: 
a) the complex Fermiology, beyond the single band model; 
b) the formation of 
charge density waves and polaronic Wigner crystals involving the 
electronic component in n-th Fermi surfaces beyond the Migdal approximation $\omega_0 /En_F \sim1$;
c) the strong electronic correlations in the electron fluid with very low Fermi energy, beyond the usually assumed Fermi gas approximation; 
e) the anomalous or negative dielectric response controlling the electron-electron interaction;
f) the complex inhomogeneous spatial geometry in systems with nanoscale phase separation \cite{hyper} induced by electron-lattice interaction and lattice misfit strain \cite{strain} which can give also insulator to superconductor phase transitions \cite{gin}.

\begin{figure}
\includegraphics[height=8.0cm,width=9.0cm]{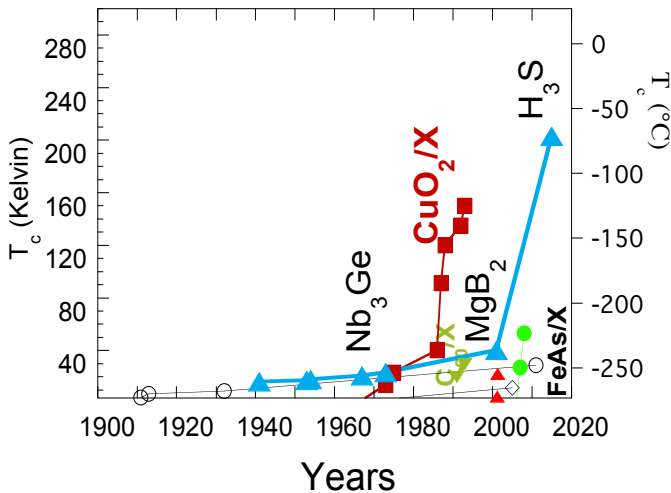}
\caption{(Color online) The superconducting critical temperature of superconducting  elements, open black circles, 
binary intermetallics, filled blue triangles, and ternary and quaternary oxide layered perovskites made of 
CuO$_2$-atomic layers with different X spacer layers (filled red squares) coefficient as a function of the 
year of the discovery.}
\label{Tcyears}
\end{figure}

 \section{Superconductivity in binary intermetallics: A15 and diborides }
 
The overall features of superconductivity in H$_3$S show that it is like 
other binary intermetallics, as diborides and A15 systems therefore 
we shown in figure 2 the evolution of the maximum critical temperature of diatomic intermetallics.

A15 compounds (like Nb$_3$Ge) have the same $Im\overline{3}m$ lattice structure 
\cite{a151} as H$_3$S above 100 GPa  (shown in Fig.1) which is made of two intertwined 
networks of atomic wires running in 3D as noticed by Friedel \cite{a152,a159}  
within a 3D density of states  \cite{arb,kle,jjp} and superconductivity appears 
in highly inhomogeneous phases at the edge of a lattice catastrophe due 
incipient structural phase transitions \cite{a153,a154,a155} giving complex pattern of defects and local lattice fluctuations \cite{a156,a157,a158}. 

Following the discovery of superconductivity in MgB$_2$ in 2001 it was 
first proposed \cite{mgb0} the breakdown 
 of the Eliashberg theory  and the need  of BPV theory including shape resonance
 to describe this system made of multiple condensates and a single critical temperature.
 The breakdown of the Eliashberg theory was rapidly accepted by the community. 
 The theoretical assumptions used by standard BCS theory, considering a single band and single gap, fail due to anisotropic
  pairing in the clean limit. The standard BCS formulas failed to predict accurately the unusually 
  high transition temperature, the effects of isotope substitution on the critical transition temperature, 
  and the anomalous specific heat of MgB$_2$. While some authors proposed the multi-band 
  superconductivity theory  \cite{smw} in the clean limit and weak 
  coupling, it was rapidly shown the need of strong coupling anisotropic two band Eliashberg 
  theory  to describe this unexpected novel superconducting phase \cite{mgb1,mgb2}. 
  The isostructural AlB$_2$ system with only $\pi$ electrons at the Fermi level 
  becomes superconductor only when the chemical doping pushes the top of the $\sigma$ 
  band above the chemical potential giving a Lifshitz transition for the appearing of two small 
  hole-like $\sigma$ Fermi pockets\cite{mgb3}. Moreover because of zero point atomic fluctuations\cite{mgb4,boeri},
 the Fermi energy in the $\sigma$  bands is time and space dependent therefore the Migdal approximation
 $\omega_0 /E_F <1$, is violated and the Eliashberg theory fails, when  the Fermi energy 
 is tuned over a large energy range of 600 meV above the bottom of band edge.\cite{mgb5,mgb6}
Therefore in spite of the conventional phonon mediated pairing, the 40 K superconducting phase in magnesium 
diboride needs a non standard BCS theory. 
Considering all data collected in diborides doped with Sc and Al for Mg or with C for B it was possible 
to give a theoretical interpretation of the variation of the two superconducting gaps and the critical temperature as 
a function of the energy separation between the top of the $\sigma$ band and
 the chemical potential using the BPV theory\cite{mgb7}. 
 The superconducting $\sigma$ gaps are much larger than the gaps in the large  $\pi$ Fermi 
 surfaces and evolve as a function of the energy separation of the top of the $\sigma$ band 
 and the chemical potential\cite{mgb7}. The small percentage of the partial electronic
density of states, DOS($\sigma$), relative to the total DOS($tot$), usually entering in the standard 
McMillan formula for T$_c$,  gives a large superconducting $\sigma$ gap, 
which drives the full system to the highest critical temperature known in binary
 intermetallics, before the recent discovery of Eremets's group \cite{droz2}.
In a single band metal, where $\omega_0 /E_F \sim1$ the Migdal 
and the adiabatic approximations \cite{boeri} fail, the system enters
 in the BCS-BEC  crossover \cite{perali,levin} and both weak 
 coupling BCS \cite{BCS} and strong coupling Eliashberg theories breakdown. 
 On the contrary the superconductivity phase in a system with multiple condensates,
  where only one or two of the Fermi surface spots are in the  BCS-BEC 
 ctossover while other condensates in the large Fermi surfaces 
 are in the BCS regime, can be described by the  theory of Bianconi Perali Valletta (BPV) \cite{leg,mgb7} 
 correctly describing the multi-gap superconductivity at the BCS-BEC crossover
 avoiding standard approximations. The condensation energy of the pairs 
 is determined by many body configuration interaction between pairs forming 
 the single macroscopic superconducting quantum coherent phase. 
 The condensation energy \cite{ann} is relevant a) in the BCS-BEC 
 crossover $\Delta_n /E_{Fn}\sim1$ and in the formation of the superconducting quantum 
 coherent phase with a single T$_c$ in a system made of multiple gaps $\Delta_n$ 
 in different bands with different Fermi energies $E_{Fn}$. 
 The shift of the chemical potential going from the normal to the superconducting phase
  below T$_c$, becomes a key term in MBBC, while on the contrary it is considered to 
  be negligible in standard BCS. The results of the BPV theory applied to doped 
  magnesium diboride  \cite{mgb7} show that it is in a regime of multiple condensates 
  with different symmetry with the key role of the shape resonance in the superconducting 
  gaps between one condensate in the BCS-BEC crossover, in the $\sigma$ band, and 
  other condensates  in the BCS regime, in the $\pi$ bands, with an essential Josephson-like pair exchange term.

\begin{figure}
\includegraphics[height=8.0cm,width=9.0cm]{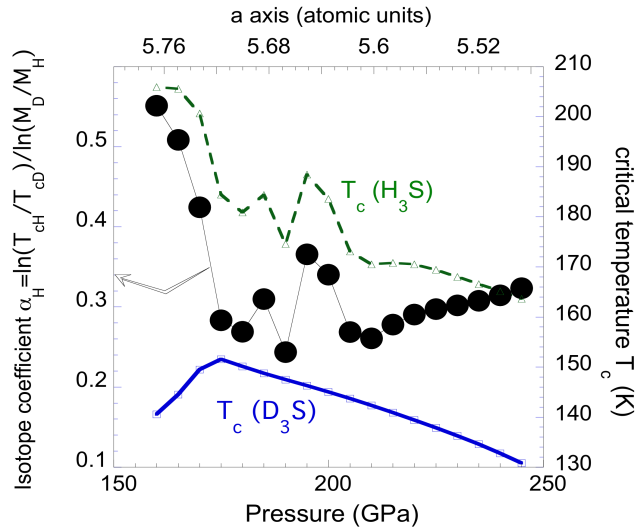}
\caption{(Color online) The pressure dependent isotope coefficient as a function of pressure 
(filled circles) calculated by interpolation of the experimental data of the critical temperature of 
H$_3$S (dashed green line) and D$_3$S) solid blue line reported by Droz et al. \cite{droz2} }
\label{isotope}
\end{figure}

\section{Lifshitz transitions.}

The topological changes of the Fermi surface of the normal phase caused by the lattice strain (as a response to pressure),
misfit strain, chemical substitution, variation of the electronic density) are called Lifshitz 
transitions \cite{Lifshitz}. In many cases Lifshitz transitions appear on points or lines of high symmetry in the Brillouin zone and are associated 
with changes of the symmetry or dimensionality of the wave functions of the electrons at the Fermi level 
in critical Fermi surfaces.
The Lifshitz transitions are revealed experimentally by anomalies in lattice parameters, in the density of states 
near the Fermi energy, in elastic properties, in anomalous thermodynamic and transport 
properties of metallic materials \cite{Varlamov}.
Lifshitz \cite{Lifshitz}. noted that at zero temperature, T = 0, Lifshitz transitions are true phase transitions of order 2.5 
(as in Ehrenfest's classification) therefore are called "2.5 Lifshitz transition". A sharp Lifshitz transition 
at T=0, at high temperature shows a crossover character. In presence of strong interactions the 2.5 phase transition 
becomes first order with a phase separation between two phases where the chemical potential in each phase is shifted 
above and below the Lifshitz transition respectively\cite{mgb6,kugel1,kugel2} driving the system
 well tuned at a Lifshitz transition on the verge of a lattice catastrophe as shown in the case of A15 and diborides.
While in the early times the interest was focused on Lifshitz transitions in single band metals, now the interest 
is addressed to Lifshitz transitions in multi-band metals with different Fermi surfaces showing multi-gap 
superconductivity. 
The shape resonance mechanism considers the relevant contribution, near the Lifshitz transition, of the exchange interaction 
between pairs in the hot spots and in all other points of the Fermi 
surfaces in the k space. We have learn that the "Devil is in the details", in fact for the optimization of a high critical 
temperature it is necessary to reach particular Lifshitz transitions where a small 
number of electrons in a new appearing Fermi surface (the hot spot) are in the extreme strong 
coupling regime in the antiadiabatic $\omega_0 /E_{Fn}\sim1$ regime in the normal phase 
without the lattice catastrophe since the majority of the electron gas is in large Fermi surfaces 
well in the Midgal approximation $\omega_0 /E_{Fn}< 1$. 
The shape resonance in superconducting 
gaps is a Josephson-like term describing a contact interaction between pairs  which increases 
the critical temperature for the pairs condensation, very similar to the Feshbach resonance 
 in ultracold gases \cite{fes1,fes2} driving up  the ratio 
 between k$_B$T$_c$ and the Fermi energy up to value of 0.2.
 In the superfluid phase in the hot spot is
  in the BCS-BEC crossover $\Delta_n /E_{Fn}\sim1$ \cite{Guidini}  and the shape 
  resonances in the superconducting gaps can be optimized to amplify the 
  critical temperature to the highest value \cite{shape}.
A large zero point lattice motion is a key term in the theory of the shape resonance
 in superconducting gaps whichj is in action in the proximity of the Lifshitz transitions.
 It was proposed to play a key role in cuprates\cite{shape1,shape2,cuprates1,isotope1,isotope3,isotope4}, 
 it was verified be in action in diborides\cite{mgb7,mgb3} and it was confirmed in iron based 
 superconductors \cite{iron1,iron2,iron3,iron4,iron5,iron6,iron7,iron8,iron9,iron10}.  
We discuss below the failure of the standard Midgal approximation and of the breakdown 
of Eliasberg theory in  H$_3$S. The large electron-phonon coupling and the high frequencies 
of H-phonons contribute to rise T$_c$ \cite{duan,papa,ari} but we claim the need 
of theories beyond Eliasberg theory. like the multigap BCS-BEC crossover (MBBC) 
theory and the BPV shape resonance mechanism to describe the gaps in the new 
small Fermi surfaces appearing at the Lifshitz transitions controlled by pressure and their 
configuration interaction with the gap in the large Fermi surface.

\section{Isotope effect in  H$_3$S.}

The isotope effect in H$_3$S \cite{droz} has provided a direct evidence of the 
involvement of the lattice degree of freedom in the pairing process. 
Therefore the isotope effect  in sulfur hydrate at high pressure H$_3$S  
has been interpreted as ruling out theories of unconventional superconductivity 
(based only on spin liquid models or magnetic interactions) and supporting 
conventional theories of superconductivity based on the role of lattice 
fluctuations. However the standard BCS theory predicts a pressure 
independent isotope coefficient 0.5 while in non standard BCS theories, 
like in the multigap anisotropic BCS,  the isotope coefficient deviates from 0.5 
as in magnesium diboride where it is 0.26 \cite{bud}. 
From the new results reported by Droz et al. \cite{droz2} we
 have extracted the isotope coefficient  as a function of pressure shown in figure \ref{isotope}. 

We can see in  figure \ref{isotope} large variations of the isotope coefficient 
reaching a minimum of 0.2 and first maximum reaching 0.5 at 170 GPa 
and  a second peak at 240 GPa. The anomalous pressure dependent isotope coefficient 
has been found in cuprates superconductors as a function of doping \cite{isotope1,isotope3,isotope4} 
with anomalies where the chemical potential crosses Lifshitz transitions driven by pressure. 
Therefore the data in figure \ref{isotope} indicate the possible presence of Lifshitz 
transitions in the pressure range showing near room temperature superconductivity.

\section{Band structure calculation of H$_3$S as a function of pressure}

\begin{figure}
\includegraphics[height=8.0cm,width=9.0cm]{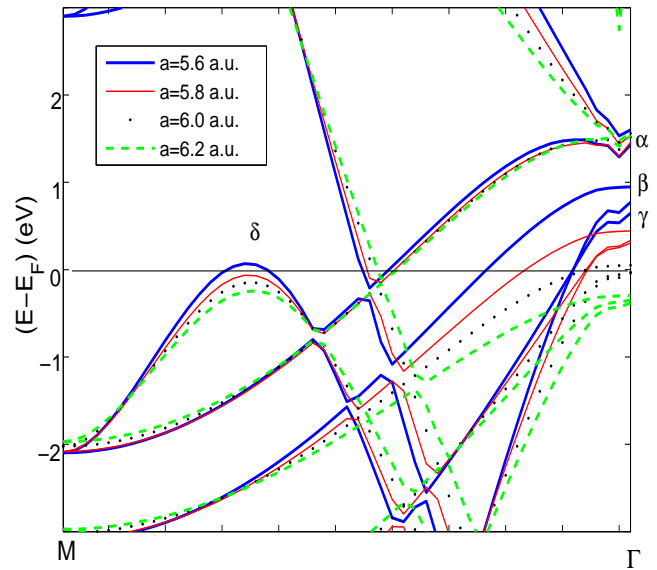}
\caption{(Color online) The dispersion of the bands in the electronic structure of H$_3$S calculated using the simple cubic cell, used for A15 compounds with $Im\bar{3}m$ structure, between $M$ and $\Gamma$ points. The band dispersion in this direction gives the narrow DOS peak near the Fermi energy.  Moreover we show the variation of this bands at different pressure with changing the lattice parameter between $a$=6.2 a.u and $a$=5.6. We show that there are bands ($\beta$ and $\gamma$ ) forming 3 small Fermi surfaces at the $\Gamma$ point and another band crosses the zero energy at the point indicated by $\delta$ which gives a new small hole pocket at high pressure.}
\label{fig5}
\end{figure}

We have performed preliminary band structure calculations\cite{noi}  of  H$_3$S  with $Im\bar{3}m$ lattice symmetry 
made using the linear muffin-tin orbital (LMTO) method \cite{lmto,bdj} and the
local spin-density approximation (LSDA) \cite{lsda}. We show in Fig. \ref{fig5} the dispersion of the bands crossing the chemical potential in the $\Gamma-M$  direction.

\begin{figure}
\includegraphics[height=8.0cm,width=9.0cm]{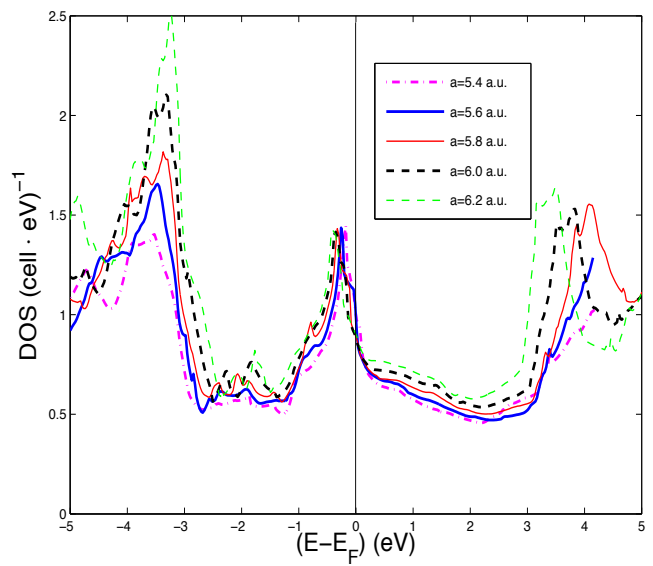}
\caption{(Color online) The total DOS for H$_3$S at different lattice constants. A sharp peak of the DOS crosses the chemical potential at high pressure.}
\label{fig1}
\end{figure}

 Self-consistent paramagnetic calculations
are made for a simple cubic unit cell containing 8 sites totally, 
used for A15 compounds with the same crystalline structure.
The details of the method have been published earlier 
\cite{jmp,ce,mgj,jb}.
The present calculations are in good
agreement with previous band structure 
calculations \cite{papa,errea,bernstein, durajski,ari}.
We confirm the presence of a narrow peak of the occupied total Density of States (DOS) very close to the chemical potential at 200 GPa i.e., for a=5.6 a.u. 
The narrow peak in the total DOS in a narrow energy range around the chemical potential is shown in Fig. \ref{fig1}. While previous papers have stated that this peak is pinned at the zero energy, we show in Fig. \ref{fig1} that the energy position of this peak relative to the chemical potential shifts with pressure. 
The narrow peak of the DOS  is pushed toward high energy by pressure and it crosses the chemical 
potential at the highest $P$ for the lattice constant smaller than a=5.8 a.u..

The narrow peak of the DOS peak is related with the flat dispersion of bands in the $\Gamma-M$  direction in the energy range of 2 eV below the chemical potential.
shown in Fig. \ref{fig5}. 
The band structure shows a first steep band with very large energy dispersion with its band edges at about 20 eV away from the chemical potential.
This first band gives the large Fermi surface shown in \cite{sanna}.
There are three other small Fermi surface pockets centered around the $\Gamma$ point 
shown in Fig. \ref{fig5} for different lattice parameter $a$.

The tops of the 
hole-like bands near the $\Gamma$ point are below $E_F$
at low pressure $P$ but they move above the chemical potential above 130 GPa pressure as shown in Fig.\ref{fig5}.
The edges of these bands cross the chemical potential as function of $P$ giving topological 
Lifshitz transitions with the appearing of new Fermi surfaces.
Moreover there is a small Fermi surface hole pocket due to the crossing of one band    
at about 2/3 of the $\Gamma-M$ distance  indicated by $\delta$  in Fig. \ref{fig5} which appears 
only for lattice parameters smaller than $a$ $\sim$ 5.8. a.u.,
i.e. in the pressure  range of highest $P$ when $T_c$ is highest. 
The top of this band goes from -0.2 eV below the chemical potential
 to +0.1 eV  when  $a$ decreases from 6.2 to 5.6 a.u..  
Finally we associate the crossing of the chemical potential by this band  at $\delta$  in Fig. \ref{fig5} with the Fermi level crossing of the narrow DOS peak in Fig. \ref{fig1}. 

Let us now consider the fact that because of the zero-point motion (ZPM) the different Fermi energies in the small pockets show large energy fluctuations like
in magnesium diboride. 
The ZPM is large because of the small mass of H atoms and it has large effects 
on the fluctuations of electronic structure. Such effects have been shown 
to be important in several different materials, even if their atomic masses are larger \cite{fesi,bron,cevib}.
Lattice fluctuations can perturb spin waves and phonons in high-T$_c$ cuprates and it 
cannot be neglected  if superconductivity relies on few phonons coupled with particular bands \cite{eri14}.
The Debye temperature for H phonons is high ($\sim 1800 K$) and the amplitude of lattice fluctuations from 
ZPM is large already at low $T$. With a force constant
$K = M \omega^2$ of 7 eV/\AA$^2$ we obtain an average amplitude $u$ of the order 0.15 \AA. 
 As seen in Fig. \ref{fig1}, the first high lying valence band has a width is about 2 Ry. This makes the band dispersion
and Fermi velocities high in the first band forming the Fermi surface centered at the R point.
The effects of energy band broadening is negligible here since the chemical potential 
is far from band edges. 
The large effect of the zero point motion is on the small Fermi pockets near the $\Gamma$  point.
The low-T energy band fluctuations in materials with narrower band widths has been found to 
be about 20 meV for $u$ in the range 0.03-0.04 \AA \cite{fesi,bron,eri14}.
From an extrapolation of these values to the conditions in H$_3$S 
we estimate that the band energy fluctuation can be of the order of 160 meV for H-bands.
 Therefore when the chemical 
potential is tuned by pressure 
near a Lifshitz transition, so that the $En_F$ in one of the 
bands is of the order of 160 meV the topology of 
the small Fermi surfaces made of small hole- or 
electron-pockets shows strong dynamical fluctuations controlled 
by the zero point lattice fluctuations.

\begin{figure}
\includegraphics[height=10.0cm,width=9.0cm]{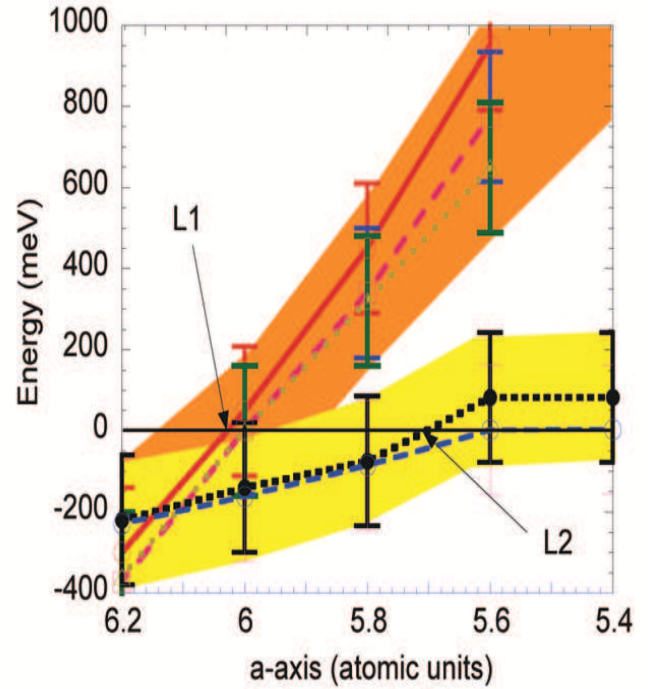}
\caption{(Color online) Energy shift
of the tops of the three hole-like bands at the $\Gamma$ point 
(solid red line, dashed red line and dotted 
 yellow line)  crossing the chemical potential at 100 GPa, 
 where the errors bars and the red region indicates the zero point energy fluctuations of the band edges. 
The filled black dots and the dotted black line show 
the energy shift of the top of the band  
crossing the chemical potential  at  about 2/3 
of the $\Gamma-M$ distance of the Brillouin zone. 
The open circles and the blue dashed line show the energy shift of the 
narrow peak of DOS near the chemical potential}
\label{fig_liftshitz1}
\end{figure}

Figure \ref{fig_liftshitz1} shows the energy position of the top of the 3 bands 
at the  $\Gamma$ point  and the band at  about 2/3 of the $\Gamma-M$ distance as a function of pressure. 
The energy spread due to zero point lattice fluctuations is
indicated by the red area for states at the $\Gamma$ point  by the yellow area for the states at  $\Gamma-M$ point. 
The shift of the 
energy position of the narrow peak in the DOS near the chemical potential 
tracks the top of the band at the  $\Gamma$-$M$ point
indicates that the chemical potential here is tuned at the Lifshitz 
transition appearing around 170 GPa for the appearing of this new Fermi surface. 
Finally these results show that in superconducting pressurized sulfur hydrate metal is made of one large Fermi surface and four small pockets tuned by pressure at Lifshitz transitions like doping tune the metal in cuprates at Lifshitz transitions for the appearing new Fermi surface arcs \cite{shape1,cuprates1,isotope1}.
 
\section{BPV theory for multi-gaps superconductors at the BCS-BEC crossover}

The standard BCS theory in undergraduate courses considers a homogeneous single band metal and describes the cooper 
pairing in weak coupling limit as due to quantum exchange of a low energy phonon between two high energy electrons at 
the Fermi level in the Migdal approximation where the gap energy is much smaller than the Fermi energy and the gap to 
critical temperature ratio is equal to 1.75. 
Non standard BCS theories diverge substantially from the standard BCS theory. Let us simplify a long story by classify 7 
different non standard BCS theories. Let us start with the first group  considering a single effective band, in a  
dirty limit, giving a single condensate: 1A) 
the standard strong coupling Eliashberg theories 
considering a high energy phonon or vibron (a quantum of intramolecular vibration) interacting with electrons having a large Fermi energy  $\omega_0 /E_F <<1$. 
Most of previous theories for high  $T_c$ in sulfur 
hydrates have proposed this approach for a large 
phonon frequency and strong coupling; 1B) the 
theories considering the case of a low density 
electron gas beyond the Midgal approximation  
 $\omega_0 /E_{Fn} >1$  where the single condensate
 is formed by BEC condensation or by BCS-BEC crossover 
  $\Delta_n /E_F\sim1$; 1C) non standard BCS theories which consider pairing mediated by electronic excitations, called exciton theories.

The second  group of non standard BCS theories considers the anisotropic BCS theory, in the clean limit, therefore 
they focus on multi-gap superconductivity where multiple gaps are formed in different in different spots of the k-space.
These non standard BCS theories consider  metals 
with multiple bands with different symmetry crossing the
 chemical potential: 2A) The multiband BCS model where 
 all multiple Fermi surfaces are in the BCS limit; 2B)
  the Fermi-Bose model where the Fermi level of a 
  first band is degenerate with a single level, occupied by
   paired electrons. Here a BCS condensate coexist with 
   bosons which undergoes a Bose condensation; 2C) 
  the extreme case of superconductivity driven mostly by exchange like interband pairing with very weak, or no intraband pairing; 2D) The theory of $shape resonance$ in superconducting gaps for multi-condensates with a small interband pairing, emerging in metals where: a first small charge density in small Fermi surfaces form condensates in the BEC- BCS crossover, beyond Migdal approximation, which coexists with a majority of charges in large Fermi surfaces forming BCS condensates.

A third group of theories consider the case of systems with relevant electronic and lattice inhomogeneity with insulator to superconductor transitions in presence of nanoscale phase separation where the Josephson-like  interaction between localized condensates play a key role.

Here we focus on the BPV (Bianconi, Perali, Valletta) theory which correctly inculde the $shape$  $resonance$ (2D) mechanism, proposed  for cuprates,  \cite{shape1,shape2,shape,cuprates1,isotope1,isotope3,isotope4}  diborides  \cite{mgb3,mgb7} and iron based superconductors  \cite{iron1,iron2,iron3}.
This mechanism (2D) is  proposed here for sulfur hydrides at very high pressure following the reported evidence for Lifshitz transitions driven by high pressure.

We recall bellow the BPV theory for multi gap superconductors at the BCS-BEC crossover including shape resonances. Let us consider a system made of multiple bands with index n, having a steep free electron like dispersion in the x direction and a flat band-like dispersion in the y direction.
The energy separation between the chemical potential 
and the bottom of the n-th band defines the Fermi energy of the n-th band. This formulation was proposed for complex systems in presence of one-dimensional lattice 
modulation or one-dimensional charge density waves where
 the chemical potential is tuned near a Lifshitz transition like in magnesium
  diborides, A15, cuprates, iron based superconductors and we propose here for sulfur hydrate at high pressure.

The formula for the superconducting critical temperature  $T_c$ in a anisotronic multi-gap
 non standard BCS scheme is given by the linearized 
 BCS equation considering the simplest case of a two dimensional system \cite{isotope1} but the extension to three dimensional system \cite{mgb7} is trivial.

\begin{equation}
\Delta_{n,k_y}=
-\frac{1}{2N}\sum_{n',\mathbf{k'}}V_{\mathbf{k},\mathbf{k}'}^{n,n'}
\frac{\tanh(\frac{E_{n,k_y}+\epsilon_{k_x}-\mu}{2T_c})}{E_{n,k_y}+\epsilon_{k_x}-\mu}\Delta_{n',k'_y},
\end{equation}
where the energy dispersion is measured with respect to the chemical 
potential. 

We consider a superconductor with multiple gaps $\Delta_{n,k_y}$ in multiple bands n with flat band-like dispersion in the y direction and steep free-electron-like dispersion in the x direction for a simple model of a two dimensional metal with a one-dimensional 
superlattice modulation in the y-direction. The self consistent equation for the gaps at
($T=0$) where each gap depends on the other gaps is given by

\begin{equation}\begin{split}
\Delta_{n,k_{y}}=-\frac{1}{2N}\sum_{n',k'_y,k'_x}
\frac{V_{\mathbf{k},\mathbf{k}'}^{n,n'}\Delta_{n',k'_y}}{
\sqrt{(E_{n',k'_y}+\epsilon_{k'_x}-\mu)^2+\Delta_{n',k'_y}^2}},
\end{split}\end{equation}

where N is the total number of wave-vectors in the discrete summation, 
$\mu$ is the chemical potential, $V^{n,n'}_{\mathbf{k},\mathbf{k}'}$ is the effective 
pairing interaction 

\begin{equation}\begin{split}
&V_{\mathbf{k},\mathbf{k}'}^{n,n'}=
\widetilde{V}_{\mathbf{k},\mathbf{k}'}^{n,n'}\\&\times\theta(\omega_0-|E_{n,k_y}+\epsilon_{k_x}-\mu|)\theta(\omega_0-|E_{n',k'_y}+\epsilon_{k'_x}-\mu|)
\end{split}\end{equation}

Here we take account of the interference effects between the 
wave functions of the pairing electrons in the different
bands, where $n$ and $n'$are the band indexes, $k_y(k_y')$ is the 
superlattice wave-vector and $k_x(k_x')$ is the component of the
wave-vector in the free-electron-like direction of the initial (final) state in the 
pairing process.

\begin{equation}\begin{split}
&\widetilde{V}_{\mathbf{k},\mathbf{k}'}^{n,n'}=-\frac{\lambda_{n,n'}}{N_0}S\\&\times\int_{S}
\psi_{n',-k'_y}(y)\psi_{n,-k_y}(y)
\psi_{n,k_y}(y)\psi_{n',k'_y}(y)dxdy,
\end{split}\end{equation}

Here $N_0$ is the DOS at $E_F$ without the lattice modulation, $\lambda_{n,n'}$ is 
the dimensionless coupling parameter, $S=L_xL_y$ is the
surface of the plane and $\psi_{n,k_y}(y)$ are the eigenfunctions in the 1D
superlattice. The gap equation need to  be 
solved iteratively. The anisotropic gaps dependent on 
the band index and on the
superlattice wave-vector $k_y$. According with Leggett \cite{leg} 
the ground-state BCS wave function corresponds to an ensemble
of overlapping Cooper pairs at weak coupling (BCS regime) and evolves to 
molecular (non-overlapping) pairs with bosonic
character and this approach remains valid also if a particular band is in the BCS-BEC crossover and beyond Migdal approximation because all other bands are in the BCS regime and in the Migdal approximation.

However in this anomalous regime, where Eliashberg theory breakdown, on density:
by increasing coupling or decreasing the density by approaching the band edge, the chemical potential $\mu$ 
results strongly renormalized with respect to the
Fermi energy $E_F$ of the non interacting system, and approaches minus half 
of the molecular binding energy of the
corresponding two-body problem in the vacuum.
 In the case of a Lifshitz transition, 
described in this paper, all electrons in the new appearing Fermi surface condense 
forming a condensate in the BCS-BEC crossover. 
Therefore at any chosen value of the
charge density for  a number of the occupied bands $N_b$, the chemical potential in the
 superconducting phase should be renormalized 
by the gap opening at any chosen value of the
charge density $\rho$ using the following formula:

\begin{equation}\begin{split}
\rho &=\frac{1}{L_xL_y}\sum_{n}^{N_b}\sum_{k_x,k_y}\left[1-
\frac{E_{n,k_y}+\epsilon_{k_x}-\mu} {\sqrt{(E_{n,k_y}+\epsilon_{k_x}-\mu)^2+\Delta_{n,k_y}^2}}\right]\\&=\frac{\delta k_y}{\pi}\sum_{n=1}^{N_b}\sum_{k_y=0}^{\pi /l_p}
\int_{0}^{\epsilon_{min}}d\epsilon \frac{2N(\epsilon)}{L_x}+\int_{\epsilon_{min}}^{\epsilon_{max}}d\epsilon \frac{N(\epsilon)}{L_x}\\&\times\left(1-
\frac{E_{n,k_y}+\epsilon_{k_x}-\mu} {\sqrt{(E_{n,k_y}+\epsilon_{k_x}-\mu)^2+\Delta_{n,k_y}^2}}\right).
\end{split}\end{equation}
  
 taking  the increment in $k_y$  as $\delta k_y = 2\pi/L_y$  
 for a size of  the considered surface a$L_x$  $L_y$ and
  in the range 
 \begin{equation*}\begin{split}
&\epsilon_{min}=max\left[0,\mu - \omega_0 - E_{n,k_y} \right],\\
&\epsilon_{max}=max\left[0,\mu + \omega_0 - E_{n,k_y} \right],\\
&N(\epsilon)=\frac{L_x}{2\pi\sqrt{\frac{\epsilon}{2m}}},
\end{split}\end{equation*}

\section{Conclusion.}

In this work we have presented the breakdown of the Eliashberg
 theory for  H$_3$S, in fact the electronic structure of sulfur hydrides 
 H$_3$S with $Im\overline{3}m$  lattice structure as function of
  pressure shows Lifshitz transitions revealed by band crossings at $E_F$ and by the shift of the narrow
peak in the density of states below the chemical potential pushed above it by lattice fluctuations associated with the hydrogen zero point motion. 
We have discussed the presence of two topological Lifshitz 
transitions at two critical pressures $P_{c1}=110$ Gpa,  and
 $P_{c2}=175$ GPa by pressure dependent electronic structure 
 calculations of H$_3$S.  At the first Lifshitz transition, 
 around $P_{c1}=110$ Gpa, three new Fermi surface spots appear at the  $\gamma$ point 
 pushed 
 up by pressure. These anisotropic
  bands are characterized by a flat dispersion in the $\Gamma-M$ 
  direction and a steep dispersion in the $\Gamma-R$ direction. 
  The second Lifshitz transition at $P_{c2}=180$ GPa is due a new 
   where the experimental critical temperature is nearly constant. 
The amplitude of the energy fluctuations of this band edge
 due to atomic zero point motion has been calculated and 
 we have found that it pushes this DOS peak above the chemical potential.
Therefore dynamical energy fluctuations of the band edge 
due to zero point motion of the hydrogen atoms is of high relevance.
 We find a colossal zero point energy fluctuation which induces a 160 meV energy
  fluctuation of the Lifshitz transitions. The present results show that the condensates
   in the 4 small hole pockets around the $\Gamma$ point and  the  small Fermi surface in the  $\Gamma-M$ direction
 are beyond the Migdal approximation $\omega_0 /E_F \sim1$, 
 including lattice dynamical zero point fluctuations. 
 In particular the condensate in the hole-like $\delta$ 
 Fermi surface pocket, associated with the sharp
  quasi-1D peak in the DOS, pushed at the chemical potential
   by pressure, is clearly in the BCS-BEC crossover regime, coexisting with other condensates in the BCS regime in other large Fermi surfaces in the pressure range where the superconducting critical temperature is near room temperature.

The emerging scenario is pairing in sulfur hydrides at high pressure in a dynamical landscape where key energy 
parameters have all the same magnitude i.e, of the order of 160 meV: 
1. the energy separation between the average position of the n$^{th}$ band edge and the chemical potential, defined 
as the n$^{th}$ Fermi energy 
controlling the "appearing or disappearing Fermi surface spot" Lifshitz transition in the Fermi surface topology,  
2. the energy separation of a peak in the Density of States and the band edge, usually controlling the "Neck opening" Lifshitz transition 
3. the amplitude of the energy fluctuation of the critical Fermi surfaces associated with zero point lattice  
fluctuations
4. The energy of the pairing interaction defining the energy of the cut-off for the formations of pairs away  
from the chemical potential

In this scenario the BCS approximations used in the standard BCS theory are no more valid and the critical 
temperature is controlled not only by the energy of the changed boson, $w_0$, and the effective electron-phonon 
coupling (given by the product of the density of states times the electron-phonon coupling constant) but also by 
the condensation energy.
While in the standard Eliashberg theory, 
the correction to the chemical potential induced by 
electron-phonon coupling is ignored on the verge 
of the Lifshitz transition this correction, which has much impact, is considered in the Multi-gaps BCS-BEC Crossover theory. 
In this situation the chemical shift from the normal to the condensed phase 
below $T_c$ is no more negligible, and the coupling should be renormalized by a factor, given by the quantum 
overlap of the condensed pairs 
 in cuprates \cite{shape1,isotope1,shape2,shape} 
 diborides \cite{mgb1,mgb5,mgb7} and iron 
 based superconductirs\cite{iron1,iron2,iron3,iron4,iron5,iron6,iron7}. 

Finally this paper shows the breakdown of Eliashberg theory
 for pressurized hydrides, supports the role of phonons 
 \cite{duan,papa,errea,bernstein,durajski,ari} but the presence 
 of Lifshitz transitions tuned by pressure need the use of the Multi-gaps BCS-BEC Crossover theory including shape resonances. 

Further work is needed to investigate i) the divergent amplitude of lattice fluctuations near the R3m to $Im\overline{3}m$ 
2nd order structural transition around 180 GPa, ii) the large mass difference between H an S which requires the 
consideration of different amplitudes of $u$ for
the two types of atoms, and it should allow for $E$ and $k$ dependences for the energy fluctuations of Lifshitz 
transitions. The electronic band calculations should be extended to large supercells needed for more precise estimates
of energy fluctuations of the electronic structure associated with spacial structural fluctuations.
We think that the discovery of superconductivity in sulfur hydrates very near room temperature has narrowed 
the number of possible road maps toward new functional superconducting materials. Further fundamental 
research on the mechanism of room temperature superconductivity in these new phase of matter are needed to 
clarify this emerging physical scenario and they will allow the definition of a protocol for the material 
design of new functional room temperature superconductors.

\end{document}